# Reconciling advective upwelling with diapycnal downwelling: A transient kinematic perspective on Munk's conundrum

Lei Han *

*China-ASEAN College of Marine Sciences, Xiamen University Malaysia, Sepang, Malaysia*
*College of Ocean and Earth Sciences, Xiamen University, Xiamen, China*

ABSTRACT

Field campaigns consistently document bottom-intensified turbulent fluxes near the seafloor, whose vertical divergence drives abyssal densification and diapycnal downwelling. This appears paradoxical under the classical assumption of steady abyssal stratification, because dense bottom waters continuously enter from Antarctic source regions. Motivated by evidence of evolving Indo-Pacific stratification, we use idealized simulations to investigate abyssal overturning circulation under transient and steady-state conditions. Using a kinematic rather than a thermodynamic water-mass-transformation approach, we diagnose advective overturning transport, isopycnal volume tendency, and kinematically diagnosed diapycnal transport in two cooling experiments with different mixing intensities, one warming experiment, and one quasi-equilibrium experiment. Advective upwelling occurs in all four experiments, whereas the direction of diagnosed diapycnal transport varies. In Type-I cooling, the isopycnal volume tendency exceeds the advective overturning in magnitude, yielding downward diagnosed diapycnal transport; in Type-II cooling, the advective overturning exceeds the isopycnal volume tendency, yielding upward diagnosed diapycnal transport. The warming and quasi-equilibrium experiments also exhibit upward diagnosed diapycnal transport. These relationships remain qualitatively consistent in potential-temperature coordinates. The Type-I cooling experiment also spontaneously produces an observationally documented bottom-intensified buoyancy-flux profile, with an e-folding length scale comparable to previous estimates. Thus, transient cooling does not uniquely determine the direction of diagnosed diapycnal transport; rather, its direction depends kinematically on the relative strengths of advective overturning and isopycnal-volume adjustment. Within the idealized framework examined here, these results provide a transient kinematic interpretation of the classical interior-downwelling conundrum while showing that diagnosed diapycnal downwelling is not an inevitable consequence of abyssal cooling.

## 1. Introduction

The global meridional overturning circulation (MOC), often referred to as the global conveyor belt or thermohaline circulation, plays a pivotal role in the Earth system by transporting heat and materials (Sarmiento and Gruber 2006; Kuhlbrodt et al., 2007; Lumpkin and Speer 2007; Talley 2013). It comprises two cells, an upper cell, with sinking in the Arctic and subarctic regions and upwelling in the Southern Ocean, and a lower cell, or abyssal cell, with sinking around the Antarctic continent and abyssal upwelling in the basin interior of the three tropical oceans (Lumpkin and Speer 2007). Each overturning cell comprises two primary components: the downwelling limb, where dense water formed in the high-latitude ocean descends to greater depths, and the upwelling limb, responsible for bringing deep waters back to the surface across the low- and mid-latitude oceans. While the downward process of "push", propelled by dense-water formation, is well-understood, the upward pathway of "pull" has long been considered a missing "crucial piece of the giant marine puzzle" (e.g., Visbeck 2007; Marshall and Speer 2012; Ferrari 2014; Voosen 2022). This study focuses on exploring the dynamics of the lower or abyssal cell.

Regarding the driving mechanism of the abyssal MOC cell, Munk and Wunsch (1998) argued from an energy perspective that "the MOC is not driven by the high latitude convective process". Rather, mixing has been identified as a pivotal driver of the abyssal MOC, particularly following the discovery of significantly elevated diffusivities near the seafloor (e.g., Toole et al., 1994; Polzin et al., 1997; Kunze et al., 2006; Lee et al.,

* Address: Bandar Sunsuria, Sepang, Selangor, Malaysia.
*E-mail address:* lei.han@xmu.edu.my.

2006; Waterman et al., 2013; Waterhouse et al., 2014; Voet et al., 2015; Mashayek et al., 2017; Kunze, 2017a, b). However, the consistent observation of bottom-intensified mixing across multiple field campaigns (e.g., Polzin et al., 1996; Morris et al., 2001; St. Laurent et al., 2001; Waterman et al., 2013; Waterhouse et al., 2014; Mashayek et al., 2017) has posed challenges to the classical advection-diffusion balance. In particular, applying this framework to bottom-intensified turbulent fluxes tends to predict diapycnal downwelling, raising questions about how such a configuration satisfies the mass balance (e.g., St. Laurent et al., 2001; Ferrari et al., 2016; Callies and Ferrari, 2018; Drake et al., 2020). This discrepancy is commonly referred to as the "interior-downwelling conundrum" or the "mixing conundrum" (Munk and Wunsch, 1998).

To address this conundrum, an influential framework has been developed that emphasizes bottom boundary layer (BBL) upwelling. This framework incorporates a turbulent buoyancy flux profile that intensifies with depth above the BBL and diminishes to zero at the seafloor within the BBL. Strong diapycnal upwelling within the BBL can then compensate for downwelling in the stratified interior (e.g., Ferrari et al., 2016; McDougall and Ferrari, 2017).

While previous theories of abyssal upwelling either assume steady stratification or neglect the displacement of abyssal isopycnals, emerging evidence suggests that the abyssal Indo-Pacific Ocean is not quiescent. A comparison based on data collected during the HMS Challenger expedition of the 1870s and modern hydrography suggests that the deep Pacific Ocean has cooled by 0.02 °C over the past century (Gebbie and Huybers, 2019). Repeated occupations of hydrographic sections in the Pacific basin have noted consistent deep ocean changes, "dispelling the notion that the deep ocean is quiescent" (Sloyan et al., 2013). Volume budgets below deep, cold isotherms within the Pacific basin "are not in steady state" (Purkey and Johnson, 2012), while recent Deep Argo and CTD measurements reveal widespread, although spatially varying, changes in deep and abyssal temperature (Johnson and Purkey, 2024).

Observed displacement rates of abyssal isopycnals and isotherms can reach several meters per year, comparable in magnitude to the Eulerian upwelling velocity required to balance the bottom-water formation (e.g., Purkey and Johnson, 2012; Sloyan et al., 2013; Voet et al., 2016; Purkey et al., 2019; Zhou et al., 2023). In one dynamically consistent ECCO realization exhibiting abyssal cooling in the Indo-Pacific, the associated abyssal volume tendency is approximately 17 Sv, comparable to or even greater than the abyssal volume supply from the Southern Ocean (Monkman and Jansen, 2024). These estimates suggest that isopycnal displacement and the associated volume adjustment may constitute a leading-order component of the abyssal overturning budget.

How does the relationship between advective overturning and diapycnal transport differ between transient and steady-state regimes? Here, a transient regime refers to the non-equilibrium adjustment of abyssal overturning and stratification when the processes sustaining the circulation are not in balance; local cooling and warming represent two possible manifestations of this adjustment. This question lies at the core of the present study. Addressing it will advance our understanding of the closure problem in the global thermohaline circulation. Following the diagnostic overturning decomposition developed by Han (2021), we distinguish among the Eulerian advective overturning streamfunction, the isopycnal volume tendency associated with the vertical displacement of density surfaces, and the kinematically diagnosed diapycnal transport. Its kinematic basis is consistent with established definitions of diapycnal motion (Marshall et al., 1999; St. Laurent et al., 2001; Ferrari et al., 2016), while related isopycnal-volume diagnostics have been independently applied to non-equilibrium overturning in the subpolar Atlantic and across the major ocean basins (Buckley et al., 2023; Monkman and Jansen, 2024). The decomposition has previously been applied to mechanisms of overturning variability in the Indian and Atlantic Oceans (Han, 2021, Han, 2023a,b, 2025), and some conclusions drawn from these applications have subsequently received independent support from observation-based studies (e.g., Volkov et al., 2024; Fraser et al., 2025).

To examine this question, we conduct four idealized numerical experiments of abyssal overturning: two transient cooling experiments with different mixing intensities, one transient warming experiment, and one quasi-equilibrium experiment. The experiments build on the setup of Ferrari et al., (2016), with two key modifications. First, sea surface temperature (SST) relaxation is removed from the transient experiments, so the abyssal isopycnals do not outcrop at the ocean surface. Second, a bottom-intensified buoyancy flux profile is no longer imposed, as such a constraint on the interior structure would otherwise force a cooling ocean. These modifications allow the thermal state, stratification, and buoyancy-flux structure of the interior basin to evolve dynamically through the interaction among AABW formation, mixing, and water-mass transformation. The resulting experiments are used to examine how advective overturning, isopycnal volume tendency, and diapycnal transport differ among the transient cooling, warming, and quasi-equilibrium.

This paper is organized as follows. Section 2 reviews observational evidence and reanalysis results concerning evolving abyssal stratification in the Indo-Pacific. Section 3 presents the overturning diagnostics obtained from ECCO, Section 4 describes and analyzes the idealized experiments, and Section 5 summarizes the principal findings and limitations.

## 2. Observational evidence and reanalysis data on a cooling abyss of the Indo-Pacific

### 2.1. Bottom-intensified diffusive fluxes

As mentioned earlier, field measurements since the 1990s have discovered significantly elevated diffusivities near the seafloor, particularly in areas with distinct topographic features and regions of heightened internal-wave energy (e.g., Toole et al., 1994; Kunze et al., 2006; Lee et al., 2006; Waterman et al., 2013; Waterhouse et al., 2014; Voet et al., 2015; Mashayek et al., 2017; Kunze, 2017a, b). The bottom intensification of turbulent dissipation rates or vertical diffusive fluxes has been consistently observed across various field campaigns (e.g., Polzin et al., 1996; Morris et al., 2001; St. Laurent et al., 2001; Waterman et al., 2013; Mashayek et al., 2017).

As introduced in Section 1, the bottom-intensified diffusive fluxes pose a challenge to the classical advection-diffusion balance model by resulting in interior downwelling (Munk, 1966; Munk and Wunsch, 1998). To address this conundrum or paradox, attempts have incorporated bottom-intensified diffusive flux profiles above the BBL, in line with observations (e.g., Ferrari et al., 2016, hereafter referred to as FMM16; McDougall and Ferrari, 2017). Due to bottom-enhanced mixing, the water parcel loses buoyancy and thus "becomes denser" (FMM16). The process is depicted in Fig. 1. As density variations in the low- and mid-latitude deep ocean are primarily governed by temperature (see Fig. 3), this observational fact further supports the evidence of abyssal cooling.

### 2.2. HMS challenger expedition versus WOCE

The observed bottom-intensified fluxes suggest that the bottom ocean is experiencing a cooling phase. However, within the broader context of global warming, this raises intriguing questions: Why does this cooling trend persist in the abyss, and how can it coexist with the warming observed at the upper ocean?

A recent study has provided important insights into these puzzles. Combining an ocean model with modern and palaeoceanographic data, Gebbie and Huybers (2019) argued that the deep Pacific is still adjusting to the cooling going into the Little Ice Age from 1400 to 1900 AD. This cooling trend is corroborated by temperature changes identified between the HMS Challenger expedition of the 1870s and modern hydrography, the World Ocean Circulation Experiment (WOCE) of the 1990s. Their model finds that the deep Pacific has cooled by 0.02 °C over the past century. The WOCE-Challenger cooling trend tends to increase towards the ocean bottom in the Pacific (Gebbie and Huybers, 2019).

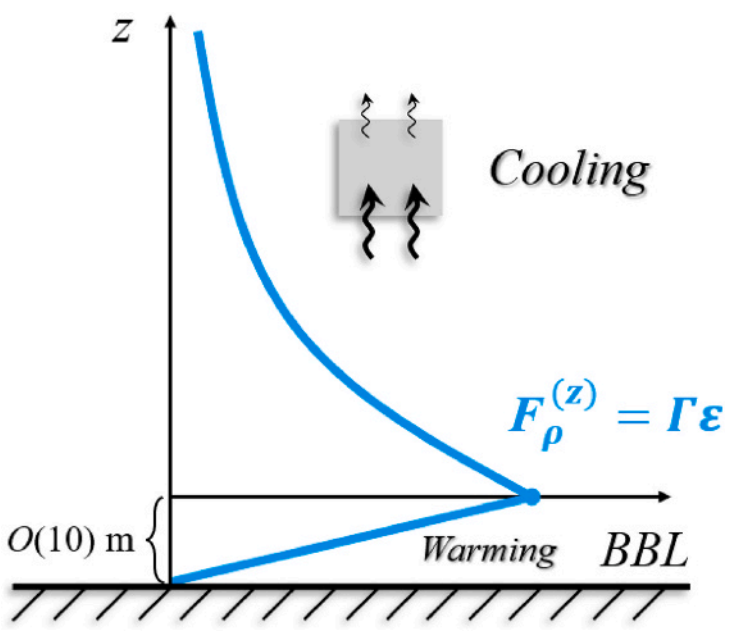

**Fig. 1.** Bottom-enhanced turbulent density flux or mixing intensity in the abyssal ocean above the BBL. $F_\rho^{(z)}$ represents vertical profile of turbulent density flux. $\Gamma$ and $\varepsilon$ denote the mixing efficiency and dissipation of turbulent kinetic energy, respectively. Curly arrows represent the turbulent density flux across the upper and lower faces of the water parcel, shown as a gray box. The size of the arrows indicates the flux intensity. A greater density influx than outflux results in density convergence within the water parcel, leading to an increase in its density. This process is equivalent to cooling if the density variation is dominated by temperature. The BBL experiences warming due to the zero-flux condition at the seafloor, but this layer is only "of order of tens of meters thick" (FMM16) in the abyssal ocean. This figure is adapted with reference to FMM16 and McDougall and Ferrari (2017).

The delayed response of the abyssal Pacific Ocean to surface cooling of the last Little Ice Age is likely associated with the formation of the Antarctic Bottom Water (AABW) and its slow creeping northward along the seafloor. Time series of surface temperature showed the cooling course in the Antarctic region began around 600 AD, peaking between 1400 and 1600 AD (Gebbie and Huybers, 2019; Johnson and Purkey, 2024). According to the water-age distribution in the deep ocean, as estimated by a data-constrained ocean circulation model (see Fig. 11 and Fig. 12a of DeVries and Primeau, 2011), the oldest water in the North Pacific is approximately 1400 years old. Based on radiocarbon observations (e.g., Sarmiento and Gruber, 2006; Matsumoto, 2007; DeVries and Primeau, 2011; Gebbie and Huybers, 2012), the cooling signal may have just reached the northern tip of the Pacific, with its tail having just passed through the Southern Ocean region. A schematized extent of this abyssal cooling signal in its northward migration is provided in Fig. 2. While uncertainties remain regarding the exact timing of the last Little Ice Age and the age of bottom water, these findings offer a plausible explanation for the cooling phenomenon currently observed in the Indo-Pacific abyss, which is consistent with the observed bottom-intensified mixing.

### 2.3. Reanalysis data capturing abyssal cooling

One of the reanalysis datasets that capture abyssal cooling is the Estimating the Circulation and Climate of the Ocean State Estimate (ECCO) (Forget et al., 2015; Fukumori et al., 2017). Previous studies have observed abyssal cooling using ECCO (Wunsch and Heimbach, 2014; Liang et al., 2015). With the latest version of ECCO (v4r4), Monkman and Jansen (2024) identified a significant shoaling trend of abyssal isopycnals throughout the Indo-Pacific region, providing support for the ongoing cooling phase. A complementary analysis of ECCO v4r4 by (Meza and Gebbie, 2025) attributed most of the simulated mid-depth North Pacific cooling to adiabatic isopycnal heave, further highlighting the importance of isopycnal displacement in a transient ocean.

ECCO is generated by best fitting the numerical simulation of an ocean general circulation model, the Massachusetts Institute of Technology general circulation model (MITgcm) to over 1 billion observations (Rousselet et al., 2020, 2022). The ECCO v4r3 utilized in this study has a nominal horizontal resolution of 1° and a depth coordinate system with 50 levels. It spans the period from 1992 to 2015 with monthly resolution. The ECCO solution is particularly well-suited for closed budget analyses due to its dynamical consistency. This dataset has also been employed in a recent study on the same question of abyssal upwelling (Wunsch, 2023).

To examine the densification trend in the abyssal Indo-Pacific, the potential density referenced to 4 km ($\sigma_4$) at a standard depth level of 4264 m is chosen for illustration. It shows the difference between two 5-year periods, 2011−2015 and 1995−1999 (Fig. 3). The selected averaging window is intended to minimize the impact of seasonal variability, as previously noted in the abyssal Indian Ocean (Han 2021). The majority of the abyssal basins in the Indo-Pacific shown in Fig. 3a are experiencing densification, primarily driven by cooling temperature (Fig. 3b, c). A linear regression analysis on the isopycnal depth of 45.9 $\sigma_4$ reveals a shoaling trend significant at the 0.01 level over the 24-year ECCO period, affecting more than 99 % of the area in the North Pacific covered by this isopycnal (see Section 3.2).

Although ECCO shows a cooling trend in the abyss, this does not contradict the repeat hydrographic observations that report deep ocean warming (e.g., Rintoul, 2007; Zenk and Morozov, 2007; Menezes et al., 2017; Purkey et al., 2019; Zhou et al., 2023). Most of these observational sections are located in the Southern Ocean, where ECCO also indicates a warming trend. As shown in Fig. 3, the field sections (black segments) exhibiting warming and freshening AABW—across the Pacific, Indian, and Atlantic sectors as well as the Vema Channel—coincide with regions where ECCO shows similar warming patterns (Fig. 3b). This consistency supports the speculation that the Southern Ocean abyss has already transitioned to a warming phase, as illustrated in Fig. 2.

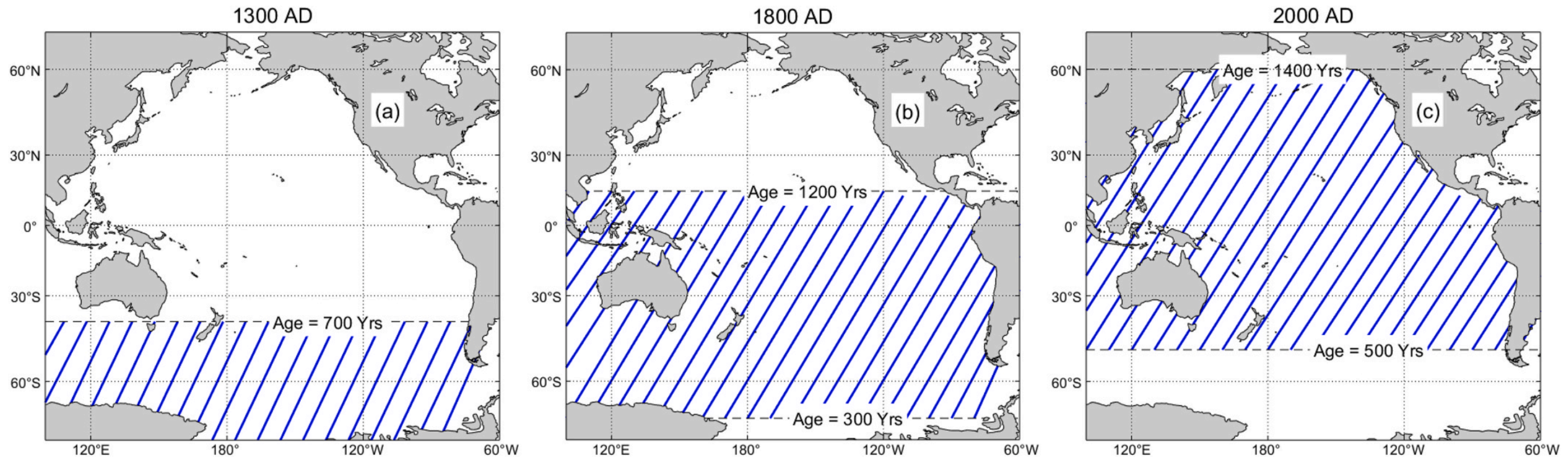

**Fig. 2.** Speculated migration of the abyssal cooling signal originating from Antarctica during the last Little Ice Age, spanning from 600 AD to 1500 AD for the Antarctic surface temperature (Gebbie and Huybers 2019). Panels (a), (b), and (c) depict the signal extent in the years 1300 AD, 1800 AD, and 2000 AD, respectively. The ages of the abyssal waters are adopted from the zonally averaged mean last-passage time of waters at a depth of 4000 m in the Pacific, based on a data-constrained ocean circulation model (DeVries and Primeau 2011).

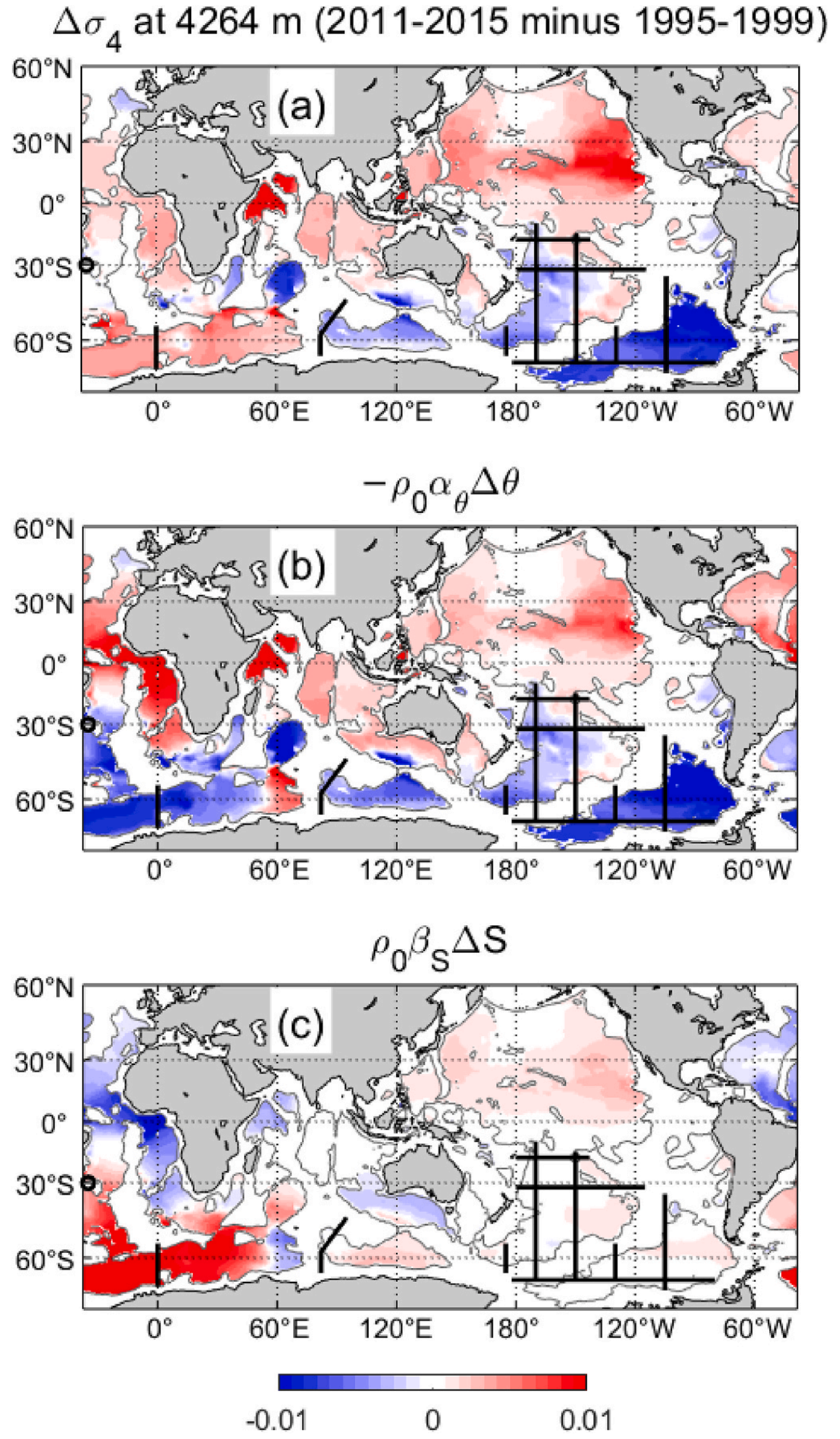

**Fig. 3.** Change in potential density (unit: kg/m$^3$) referenced to 4 km ($\sigma_4$) at a depth of 4264 m between two 5-year-period averages (2011–2015 minus 1995–1999) (a). The relative contributions from potential temperature (b) and salinity (c) to this density difference are estimated using the local thermal expansion coefficient, $\alpha_\theta$, and the saline contraction coefficient, $\beta_S$. Black solid lines/circle in each panel mark the repeat hydrographic sections/station that have revealed a warming trend of AABW around the Antarctica (see text). The warm color represents densification (a), cooling (b), or salinification (c) while the cold color represents the opposite changes. Data: ECCO v4r3.

Nevertheless, mismatches between ECCO and hydrographic data remain. For example, the freshening trend observed in hydrographic sections is not as pronounced in ECCO. More critically, there is an excessive salinification trend in the Atlantic sector of ECCO (Fig. 3c). This increase in salinity is so substantial that it outweighs the warming effect, leading to densification. This result contradicts the repeat hydrographic data anyway (Zhou et al., 2023).

Using data from Deep Argo and historical Conductivity-Temperature-Depth (CTD) instruments, Johnson and Purkey (2024) analyzed temperature trends in the abyssal ocean over the past 30 years. Even though their findings reveal a pervasive warming trend throughout the abyss, this trend weakens significantly in the Northern Hemisphere below 4,000 m, where cooling trend emerges in certain regions (see Fig. 3a in Johnson and Purkey, 2024). Furthermore, they note that the intensity of the warming trend decreases with increasing distance from the Antarctic, a spatial pattern also reflected in the ECCO dataset (Fig. 2). This distribution also aligns with the Little Ice Age hypothesis outlined in Section 2.1.

To encapsulate Section 2, although direct temperature observations leave uncertainty as to whether a cooling or warming trend prevails in the modern abyssal Indo-Pacific (Gebbie and Huybers, 2019; Johnson and Purkey, 2024) and how much this abyssal cooling signal is sourced from the last Little Ice Age, growing field measurements of bottom-intensified mixing intensity bolster the cooling hypothesis (Section 2.1). The cooling-associated mixing profile has also been prescribed in idealized model configurations used to investigate abyssal circulation under the BBL upwelling framework (e.g., Ferrari et al., 2016).

Although the fidelity with which ECCO represents observed abyssal temperature trends remains uncertain, the dataset provides a dynamically consistent realization characterized by abyssal cooling (Wunsch and Heimbach, 2014; Liang et al., 2015; Monkman and Jansen, 2024). Rather than treating the ECCO solution as definitive evidence that the contemporary Indo-Pacific abyss is cooling, we use it to present the behavior of the abyssal overturning circulation under a cooling scenario. Accordingly, the next section employs ECCO to analyze the dynamics of the meridional overturning circulation in the abyssal Indo-Pacific.

## 3. Contemporary MOC feature in the abyssal Indo-Pacific

### 3.1. Isopycnal volume tendency and kinematically diagnosed diapycnal transport

In this section, four overturning diagnostics, comprising three types—advective overturning transport, isopycnal volume tendency, and kinematically diagnosed diapycnal transport—are diagnosed from ECCO. The conventional Eulerian MOC streamfunction, written as $\psi_{Eul}$, is obtained by vertically integrating the meridional velocity from the seafloor upward across a zonal section that spans from the western boundary, $x_w$ to the eastern boundary, $x_e$,

$$\psi_{Eul}(y,z,t) = -\int_{-H}^{z} dz' \int_{x_w}^{x_e} v(x,y,z',t)dx, \tag{1}$$

where $v$ denotes the meridional Eulerian velocity component, and $H$ the maximum depth of the domain. The Eulerian MOC streamfunction can also be integrated in the density coordinate as follows,

$$\psi_\sigma(y,\sigma,t) = -\int_{-H}^{z_\sigma} dz' \int_{x_w}^{x_e} v(x,y,z',t)dx \tag{2}$$

where $z_\sigma$ represents the depth of the isopycnal σ. These two Eulerian streamfunctions are both obtained by integrating the advective velocity, and are therefore referred to as the advective MOC. They quantify the volume transport across a zonal section below a specified depth or density.

Next, we define another overturning diagnostic, the isopycnal volume tendency (written as $\psi_{iso}$), which quantifies the rate of volume change below a given isopycnal. It can be obtained by horizontally integrating the displacement velocity of isopycnals as follows (Han, 2021, Han, 2023a),

$$\psi_{iso}(y,z,t) = -\int_{y}^{y_N} dy' \int_{x_w}^{x_e} w_{iso}(x,y',z,t)dx, \tag{3}$$

where $w_{iso}$ denotes the displacement velocity of isopycnals mapped onto standard depth levels, and $y_N$ is the northern boundary of the Indo-Pacific basin, where the isopycnal volume tendency is zero by defini-tion. The isopycnal vertical velocity, $w_{iso}$, is estimated from the finite difference in isopycnal depth between two consecutive snapshots in the reanalysis data. More details on the algorithm are available in Han (2021). The isopycnal volume tendency, as defined in Eq. (3), has been applied in the Indian Ocean (Han 2021) and the Atlantic Ocean (Han, 2023a,b, 2025) to examine the mechanisms governing MOC variability. Independently, Monkman and Jansen (2024) used a closely related diagnostic, the change in volume below a given isopycnal, to investigate the non-equilibrium effect on overturning circulation across the major ocean basins.

In our previous studies, this quantity was termed the "sloshing overturning streamfunction" and denoted by $\psi_{slo}$. That terminology was inaccurate because Eq. (3) is not derived from a nondivergent velocity field and therefore does not define a conventional advective streamfunction. In the present study, the quantity is renamed the "isopycnal volume tendency" and denoted by $\psi_{iso}$, which more directly describes its physical meaning and is closely related to the volume-change diagnostic used by Monkman and Jansen (2024).

Diapycnal processes can be investigated from either thermodynamic or kinematic perspectives (Marshall et al., 1997). The thermodynamic water-mass-transformation approach diagnoses transformation rates from buoyancy sources and sinks (e.g., Walin, 1982; McDougall, 1984; Marshall et al., 1997), whereas the kinematic approach diagnoses diapycnal motion from the movement of water relative to moving density surfaces (e.g., Marshall et al., 1999; St. Laurent et al., 2001; Ferrari et al., 2016). The present study adopts the latter perspective.

In the kinematic framework, the difference between the Eulerian vertical velocity, $w_{Eul}$, and the isopycnal displacement velocity, $w_{iso}$, represents the vertical component of water motion relative to moving isopycnals. This relative velocity has been interpreted as the diapycnal velocity in previous studies (e.g., Marshall et al., 1999; St. Laurent et al., 2001; Ferrari et al., 2016) and is denoted here by $w_{dia}$:

$$w_{dia} = w_{Eul} - w_{iso}. \tag{4}$$

As the Eulerian overturning streamfunction in Eq. (1) can equivalently be expressed as the horizontal integral of $w_{Eul}$ over the basin area north of $y$, basin integration of Eq. (4) therefore gives the diagnosed diapycnal transport, $\psi_{dia}$:

$$\psi_{dia} = \psi_{Eul} - \psi_{iso}. \tag{5}$$

where

$$\psi_{dia} = -\int_{y}^{y_N} dy' \int_{x_w}^{x_e} w_{dia}(x, y', z, t) dx. \tag{6}$$

Here, $\psi_{dia}$ is interpreted as a kinematically diagnosed diapycnal transport, rather than an independently calculated thermodynamic water-mass-transformation rate. Although the two approaches describe related aspects of diapycnal processes, their general equivalence under arbitrary transient conditions is neither assumed nor demonstrated in this study.

A summary of these three types of overturning diagnostics and their associated velocity definitions is provided in Table 1. The results diagnosed from ECCO are presented in the following section.

**Table 1**
Definitions and properties of the three overturning diagnostics analyzed in this study. The related vertical velocity component is listed in the final column.

| Variable name | Symbol | Tracked object | Related vertical velocity component |
|---|---|---|---|
| Advective MOC | $\psi_{Eul}$ , $\psi_{res}$ or $\psi_{\sigma}$ | Water parcels or material surfaces | Eulerian velocity ($w_{Eul}$) or residual velocity ($w_{res}$) |
| Isopycnal volume tendency | $\psi_{iso}$ | Density or property surfaces (e.g., isopycnals) | Displacement rate of isopycnal ($w_{iso}$) |
| Kinematically diagnosed diapycnal transport | $\psi_{dia}$ | Water motion relative to moving isopycnals | Kinematically diagnosed diapycnal vertical velocity ($w_{dia}$ = $w_{Eul} - w_{iso}$) |

### 3.2. Indo-Pacific overturning diagnostics in ECCO

The four time-averaged overturning diagnostics ($\psi_{Eul}$, $\psi_{\sigma}$, $\psi_{iso}$, and $\psi_{dia}$) for the Indo-Pacific basin, computed using ECCO, are presented in Fig. 4. A large-scale anti-clockwise abyssal overturning cell is clearly visible in the Eulerian MOC ($\psi_{Eul}$), indicating significant upwelling of the bottom water towards the intermediate depths. When comparing the two streamfunctions in depth and density coordinates (Fig. 4a and b), the abyssal cells in both solutions show close agreement in extent and strength. On one hand, this consistency serves to cross-verify the results; on the other hand, it indicates that zonal tilting of isopycnals in this region is relatively small, in contrast to the situation observed for the Deacon cell in the Southern Ocean (e.g., Döös et al., 2008). Supporting our results, similar structures of the Indo-Pacific abyssal cell, in both depth and density coordinates, have also been reported in previous ECCO-based studies (Rousselet et al., 2021; Rogers et al., 2023; Monkman and Jansen 2024).

Examining the isopycnal volume tendency, $\psi_{iso}$ (Fig. 4c), reveals that the abyssal isopycnals are rising at a rate even faster than that of the water parcels (Fig. 4a). Given that temperature dominates the density change in the abyssal ocean, this phenomenon corresponds to abyssal cooling, which aligns with the observed vertical divergence of turbulent buoyancy flux associated with bottom-intensified mixing (Fig. 1). Furthermore, the difference between $w_{Eul}$ and $w_{iso}$ leads to a negative diapycnal vertical velocity, $w_{dia}$, which is manifested as a prominent downwelling overturning cell, as shown in Fig. 4d. The same qualitative result of abyssal advective upwelling coexisting with diagnosed diapycnal downwelling is obtained using the residual MOC streamfunction, which additionally accounts for parameterized eddy-induced transport (Fig. S1 of the Supplementary Materials; see Section 4.2.1 for the definition of the residual MOC streamfunction).

The basin-integrated isopycnal inflation in the abyssal Indo-Pacific has been identified by Monkman and Jansen (2024). Here the spatial pattern and significance of this isopycnal shoaling in the abyssal North Pacific is examined. Using the isopycnal 45.9 $\sigma_4$ as an example, a linear regression is applied to the isopycnal depth during the 24-year ECCO period at each grid point of the North Pacific. The results, shown in Fig. 5, reveal a shoaling trend of this isopycnal across most of the region, significant at a level of 0.01. Additionally, this analysis offers an alternative means of estimating the isopycnal displacement rate, $w_{iso}$, which is determined as the slope of the linear regression applied to the isopycnal depth over the ECCO period. According to the new results as shown in Fig. 6, the isopycnal volume tendency ($\psi_{iso}$) obtained using this method is ~2 Sv weaker than the method described previously (Fig. 4c), yet it remains stronger than the Eulerian MOC ($\psi_{Eul}$) (Fig. 6a). As a result, it still yields a notable downward kinematically diagnosed diapycnal transport (i.e., positive $\psi_{dia}$) (Fig. 6b). These results suggest that, despite uncertainties in the absolute strength of the abyssal meridional overturning, its direction and qualitative structure—particularly the downwelling diapycnal component—remain robust.

### 3.3. Discussion

The relationship among the three vertical velocities defined in Table 1—the Eulerian velocity ($w_{Eul}$), the displacement rate of isopycnal ($w_{iso}$), and the diapycnal velocity ($w_{dia}$) —has been illustrated for transient stratification scenarios in a previous study (see Fig. 4 of Han 2021). Here, these kinematic relationships are extended to include both upward and downward Eulerian motion and are illustrated schematically in Fig. 7.

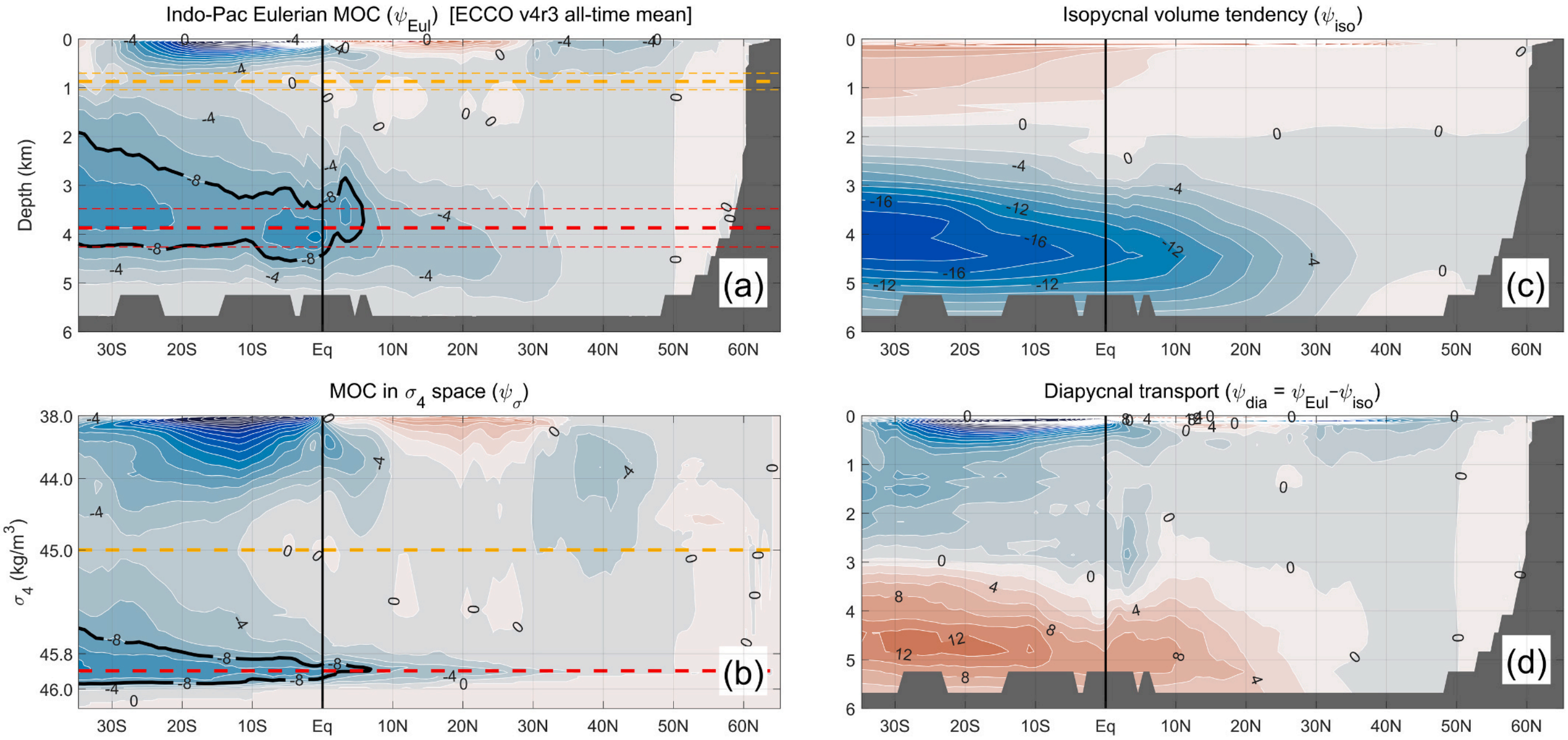

**Fig. 4.** The time-averaged overturning diagnostics in the Indo-Pacific Ocean. (a) Eulerian MOC streamfunction in the depth coordinate ($\psi_{Eul}$); (b) Eulerian streamfunction in the density coordinate of $\sigma_4$ ($\psi_\sigma$); (c) Isopycnal volume tendency ($\psi_{iso}$); (d) Kinematically diagnosed diapycnal transport ($\psi_{dia}$). (a) and (b) both belong to advective MOC streamfunctions. Positive values (warm color) denote clockwise-overturning (downwelling) cells while negative values (cold color) denote anti-clockwise (upwelling) cells. The thick/thin dashed lines in panel (a) represent the mean/s.d. depth of two isopycnals, 45.0 $\sigma_4$ and 45.9 $\sigma_4$, respectively, as indicated in panel (b). The contour of –8 Sv in the deep cell is thickened for clearer comparison in (a) and (b). The black arrows indicate the vertical direction of overturning in the abyss of interior basin. Data: ECCO v4r3. Units: Sv. Contour interval: 2 Sv.

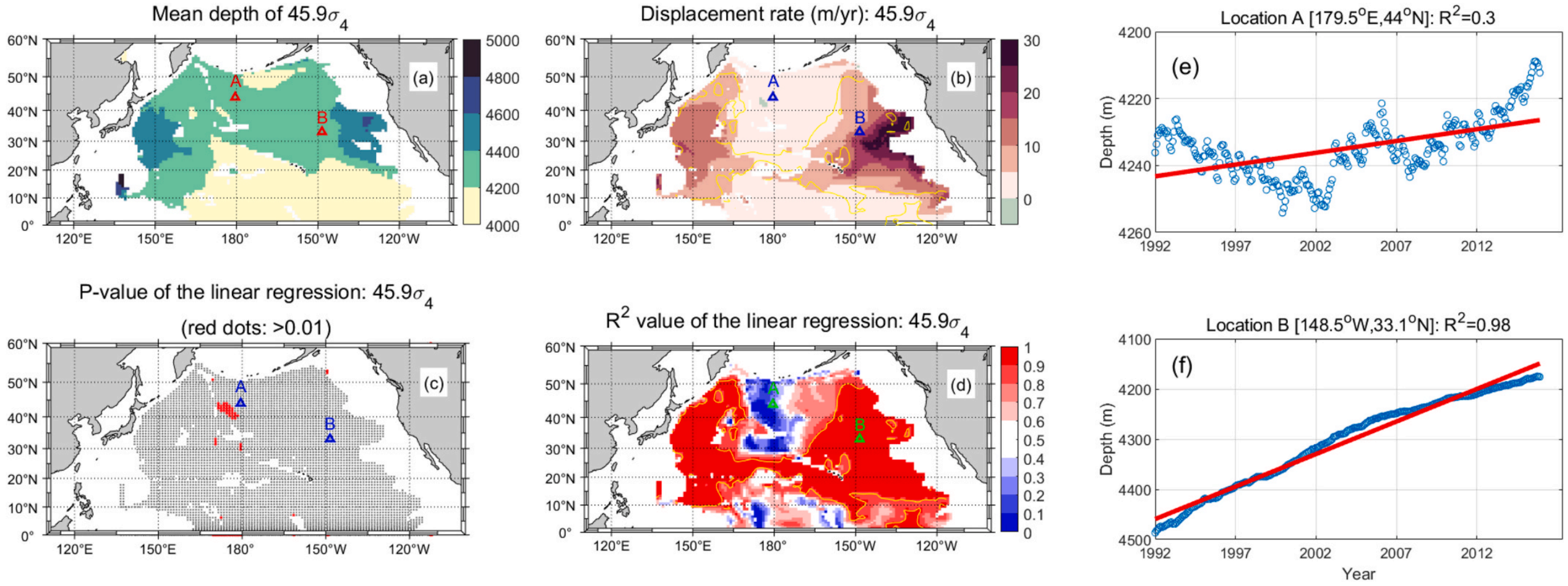

**Fig. 5.** Shifting of the abyssal isopycnal, 45.9 $\sigma_4$, in the North Pacific. (a) Mean depth of the isopycnal (m); (b) Displacement rate, obtained as the shoaling trend of isopycnal depth (m/year); (c) The p-values of the linear regressions applied to the isopycnal depth at individual grid points, with red and black dots indicating p-values greater than and less than 0.01, respectively; and (d) The coefficients of determination of the linear regression in (b) and (c). Panels (e) and (f) present examples of isopycnal shifting (blue circles) and their corresponding linear regression lines (red solid) at two representative locations, A and B, as marked in panels (a)-(d). Data: ECCO v4r3.

The theoretical basis for determining the direction of diapycnal motion is the unsteady advection–diffusion balance (e.g., McDougall 1984; Marshall et al., 1999; St. Laurent et al., 2001; de Lavergne et al., 2016; FMM16):

$$w_{dia}\frac{\partial\rho}{\partial z} = -\frac{\rho}{g}\frac{\partial}{\partial z}(\Gamma\varepsilon), \tag{7}$$

where the diffusion term on the right-hand side of Eq. (7) can be expressed equivalently as:

$$-\frac{\rho}{g}\frac{\partial}{\partial z}(\Gamma\varepsilon) = \frac{\partial}{\partial z}\left(\kappa\frac{\partial\rho}{\partial z}\right) = -\frac{\partial F^z_\rho}{\partial z} = \frac{\partial F^z_b}{\partial z}. \tag{8}$$

Here, $F^z_\rho = \overline{w'\rho'}$ and $F^z_b = \overline{w'b'}$ represent the vertical turbulent density flux and buoyancy flux, respectively. Thus, the direction of $w_{dia}$ is determined by the vertical convergence or divergence of the turbulent density or buoyancy flux.

Under a bottom-intensified diffusive flux, Eq. (7) yields a negative, or downward, diapycnal velocity, regardless of the direction of the

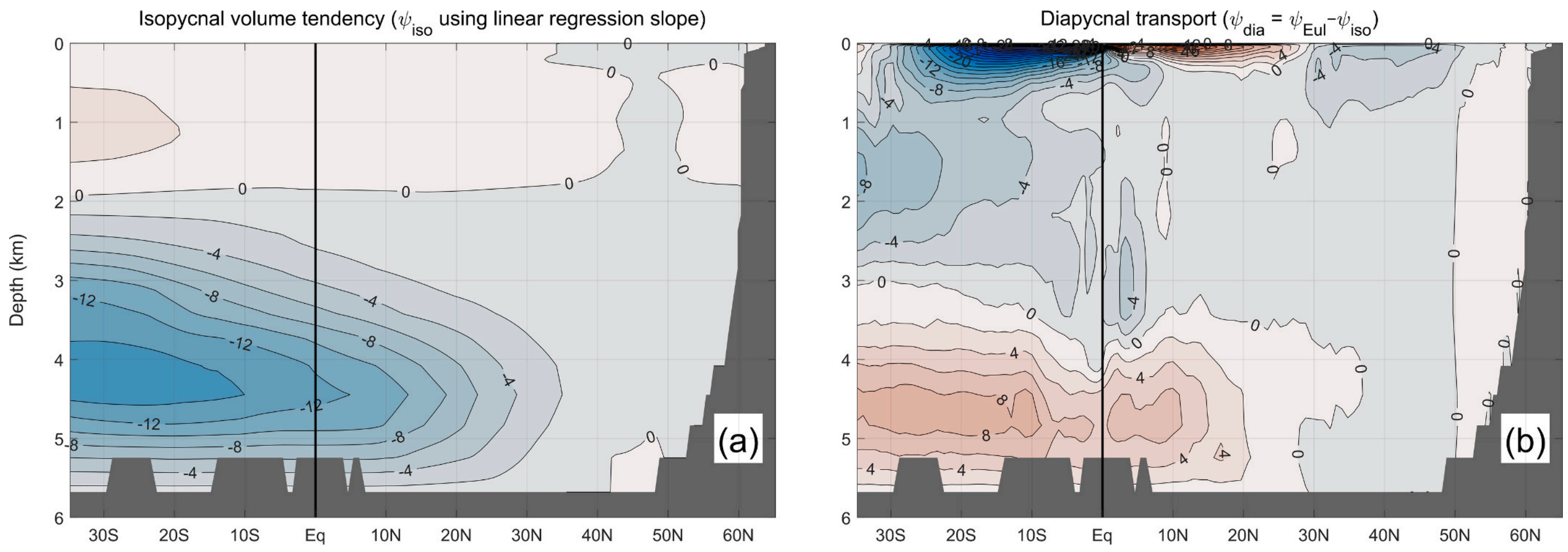

**Fig. 6.** As in Fig. 4 (c) and (d), except that the isopycnal volume tendency in (a) is calculated from the linear trend of isopycnal depths as demonstrated in Fig. 5 (a). The kinematically diagnosed diapycnal transport in (b) is obtained as the difference between Fig. 4a and the isopycnal volume tendency in (a). This alternative method yields results qualitatively consistent with those in Fig. 4. Data: ECCO v4r3.

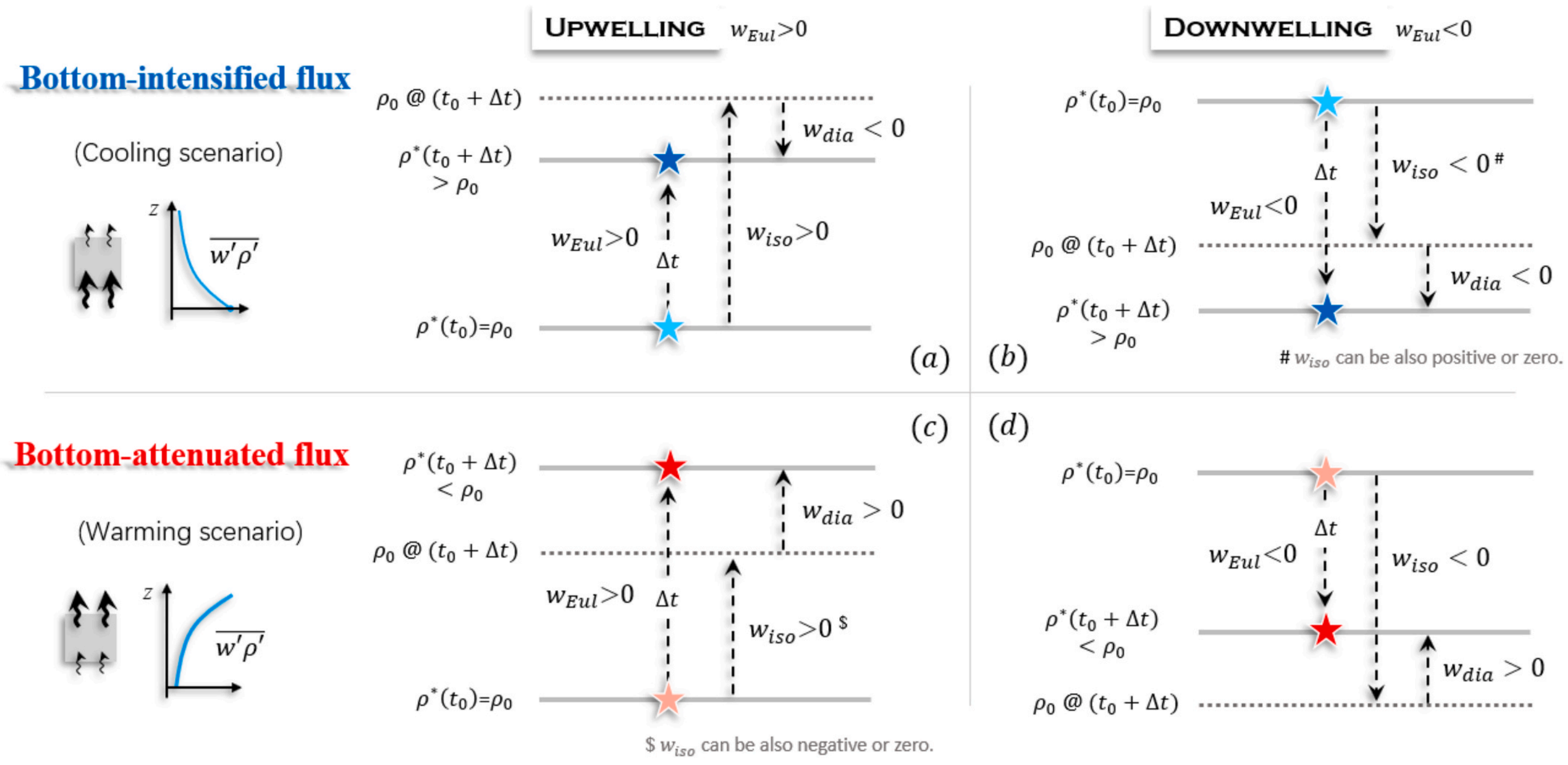

**Fig. 7.** Conceptual schematic of the kinematic relationships among the Eulerian velocity ($w_{Eul}$), isopycnal displacement velocity($w_{iso}$), and diapycnal velocity ($w_{dia}$) in the context of bottom-intensified diffusive flux (a, b) and bottom-attenuated diffusive flux (c, d). Each scenario considers both upwelling (a, c) and downwelling (b, d) of water parcels. Stars represent water parcels; dashed arrows indicate vertical velocities; grey lines show the water parcel's initial position with density $\rho_0$ at time $t_0$, and its final position with density $\rho^*$ after an interval $\Delta t$; dotted lines mark the new position of isopycnal $\rho_0$ at time $t_0+\Delta t$. Bottom-intensified and bottom-attenuated flux profiles are depicted in the left panels (curly arrows denoting the vertical density fluxes).

Eulerian motion. In this situation, the material and density surfaces move at different rates because the water parcel becomes denser and separates from its original isopycnal. This is a local relationship between water motion and density-surface motion under flux divergence; it does not imply that every basin-scale cooling state must exhibit downward diapycnal transport. Fig. 7a and b schematically illustrate these relationships for upward and downward Eulerian motion, respectively.

Conversely, under a bottom-attenuated diffusive flux that produces local warming, Eq. (7) yields a positive, or upward, diapycnal velocity for either direction of Eulerian motion. This is consistent with the upward interior diapycnal velocity obtained under prescribed bottom-attenuated fluxes in the numerical experiments by FMM16. Fig. 7c and d provide conceptual illustrations of these theoretically and numerically supported relationships.

Under steady stratification, the isopycnal displacement velocity vanishes ($w_{iso}=0$). The kinematic relationship ($w_{dia}=w_{Eul}-w_{iso}$) therefore reduces to ($w_{dia}=w_{Eul}$), meaning that the Eulerian and diapycnal vertical motions coincide. For upward Eulerian motion, this steady-state balance can be regarded as a limiting case of the warming relationship illustrated schematically in Fig. 7c. Water parcels are progressively warmed as they ascend, while the upward advective heat transport is balanced by downward diffusive heat transport, thereby maintaining stationary isopycnals. This is the balance assumed in the classical "abyssal recipes" framework (Munk, 1966; Munk and Wunsch, 1998).

Taken together, Eq. (7) and the kinematic relationship ($w_{dia}=$

$w_{Eul} - w_{iso}$) link local diapycnal motion to the relative movement of water and density surfaces. They do not establish a one-to-one mapping between the basin-mean thermal tendency and the direction of diapycnal transport. The ECCO diagnostics in Figs. 4 and 6 provide one cooling realization with downward diagnosed diapycnal transport. The idealized experiments below test a broader range of cooling, warming, and steady-state relationships.

## 4. Simulated overturning diagnostics under transient and quasi-equilibrium conditions

### 4.1. Model description

To investigate abyssal MOC behavior under warming, cooling, and quasi-equilibrium scenarios, we perform idealized box-model simulations with a periodic or reentrant channel using the MITgcm (Marshall et al., 1997). The model represents a closed basin connected to a laterally unbounded channel to the south. Related configurations have been widely employed in studies examining the mechanisms driving the global MOC (e.g., Ito and Marshall, 2008; Wolfe and Cessi, 2010; Nikurashin and Vallis, 2011; Wolfe and Cessi, 2011; Munday et al., 2013; Bell, 2015; Mashayek et al., 2015; FMM16).

Because FMM16 investigated abyssal overturning using a related model configuration, we adopt much of its setup here (Table 2), with two key modifications.

First, we relax the constraint of a prescribed bottom-intensified turbulent density flux. As discussed in Section 3, this constraint imposes a cooling in the abyss, and is therefore not suitable for simulating warming and steady conditions.

Second, the SST restoration used by FMM16 allows abyssal isopycnals to ventilate to, or outcrop at, the surface of the "Southern Ocean". However, field hydrographic sections indicate that many abyssal isopycnals do not outcrop at the ocean surface (e.g., Schlitzer, 2000; Talley et al., 2011; Rintoul, 2018; Moorman, 2019; Wunsch 2023). Unlike the formation of North Atlantic Deep Water, which is primarily driven by surface cooling, the formation of AABW involves a more complex sequence of processes. In Antarctic coastal polynyas, intense surface cooling, vigorous sea ice production, and interactions with the base of ice shelves produce supercooled, high-salinity Dense Shelf Water. This dense water mass pools on the continental shelf and subsequently cascades down the continental slope as a gravity-driven plume, entraining ambient Southern Ocean waters and eventually forming AABW (Moorman, 2019; Silvano et al., 2023). These processes effectively decouple bottom water formation from direct surface ventilation, as evidenced by WOCE transect observations (Schlitzer, 2000).

Imposing SST relaxation at the surface could therefore bias the heat budget in the unventilated part of the abyssal ocean by introducing an artificial heat flux through the sea surface. To avoid this issue, such SST forcing condition is excluded from our simulations. Additionally, we extend the model domain southward to incorporate an idealized "polar region" located south of the reentrant channel. This region acts as a reservoir for the densest waters prior to their export into the basin interior, mimicking the overflow of the densest waters from its formation zones into the open ocean. While this setup simplifies the full complexity of AABW formation, it provides a practical framework for exploring its role in driving abyssal overturning circulation within the interior basin.

The model domain consists of three distinct regions: (1) a polar region spanning 70°–60°S, representing the formation region of AABW; (2) a reentrant channel from 60°–40°S, simulating the Southern Ocean; and (3) a rectangular, flat-bottom basin extending from 40°S to 60°N, representing the Indo-Pacific Ocean (Fig. 8). The basin spans 92° in longitude with a resolution of 2° × 2°. Bathymetry is uniform at a depth of 4000 m, except for a zonal seamount approximately 1500 m high, located between the polar region and the Southern Ocean, which simulates the outflow of AABW into the Southern Ocean.

To simplify the dynamics of simulated abyssal overturning circulations, we apply minimal external forcing in our experiments. Specifically, temperatures below 2000 m at the southern wall (70°S) are restored to 1°C on a monthly basis (area indicated by red dots in Fig. 8). In contrast to the setup used in FMM16, we explore an alternative configuration that avoids surface heat flux or wind forcing, for reasons outlined earlier. Following FMM16, salinity is held constant throughout, and potential density is treated as a function of temperature only. This idealization isolates the density adjustment associated with temperature variations, which provides the dominant contribution to abyssal density changes over much of the Indo-Pacific in the ECCO realization examined in Section 2.3 and Fig. 3. The initial temperature field is horizontally uniform, decreasing exponentially from 25°C at the surface grid to 2°C at the bottom grid, with an e-folding scale of 1000 m. The configuration file for our simulation is adapted from the MITgcm tutorial "Southern Ocean Reentrant Channel Example" (accessed December 2023, https://mitgcm.readthedocs.io/en/latest/examples/reentrant_channel/reentrant_channel.html).

This numerical configuration is used to examine both transient and steady-state regimes of abyssal overturning circulation. In this study, a transient regime refers to the non-equilibrium adjustment of abyssal overturning and stratification that occurs when the processes sustaining the circulation—conceptually, the "pushing" by AABW supply and the "pulling" by mixing-induced water-mass transformation—are not in balance. Cooling and warming represent two possible manifestations of this adjustment, whereas a steady-state regime is reached when these competing processes balance and the abyssal stratification becomes statistically stationary.

The transient experiments are conducted by varying the intensity of vertical mixing, specifically of the vertical diffusivity within the interior basin. In the initial cooling experiment, moderate mixing ($\kappa$=1 × $10^{-5}$ $m^2/s$) is applied, such that the AABW intrusion rate exceeds the mixing-induced water transformation rate, resulting in abyssal cooling. Conversely, the warming experiment employs a larger vertical diffusivity ($\kappa$=1 × $10^{-4}$ $m^2$/s), such that mixing transforms dense water faster than it is replenished by AABW intrusion, causing the abyss to warm. In both experiments, diffusivity in the AABW formation region (70°–60°S) is held at a low value ($\kappa$=1 × $10^{-6}$ $m^2$/s) to minimize the transformation of bottom water prior to its export to the interior basin. Each experiment employs a vertically uniform diffusivity and is initialized from a state of rest.

**Table 2**
Model parameters for the MITgcm simulations. Most parameters are adopted from the FMM16 configuration, with modifications to diffusivity and external forcing. The initial cooling and warming experiments differ solely in the vertical diffusivity ($\kappa$) applied within the interior basin; the quasi-equilibrium experiment additionally uses equatorial SST relaxation, as described in Section 4.2.2. An additional cooling experiment is also conducted as described in Section 4.2.3.

| Model parameters (variable names in MITgcm) | Values |
|---|---|
| Coordinate | Cartesian |
| Range of domain | 70°S-60°N, 0-92° |
| Horizontal resolution (delX, delY) | 2° × 2° |
| Vertical resolution (delR) | 80 m |
| Basin depth | 4000 m |
| Boundary condition (external forcing) | Southernmost grid points below 2000 m are relaxed to 1°C on a monthly basis |
| Wind stress | None |
| Time step (deltaT) | 3000 s |
| EOS type (eosType) | Linear |
| Thermal expansion coefficient $\alpha$ (tAlpha) | $2 \times 10^{-4}$ °C$^{-1}$ |
| Haline contraction coefficient $\beta$ (sBeta) | 0 |
| GM package (eddy effect) | Enabled |
| Vertical diffusivity $\kappa$ (diffKrT) | Vertically uniform throughout the domain.<br>Polar region: $1 \times 10^{-6}$ m$^2$/s<br>Interior basin:<br>• Warming experiment: $1 \times 10^{-4}$ m$^2$/s<br>• Initial cooling experiment: $1 \times 10^{-5}$ m$^2$/s<br>• Additional cooling experiment: $3 \times 10^{-5}$ m$^2$/s |
| Horizontal viscosity (viscAh) | $2 \times 10^{5}$ m$^2$/s |
| Vertical viscosity (viscAr) | $1.2 \times 10^{-4}$ m$^2$/s |
| Bottom drag coefficient (bottomDragLinear) | $5 \times 10^{-4}$ m/s |
| Coriolis parameter at southernmost point (f0) | $-1.3667 \times 10^{-4}$ s$^{-1}$ (70°S) |
| Beta-plane parameter (beta) | $1.9524 \times 10^{-11}$ m$^{-1}$s$^{-1}$(to ensure Coriolis parameter vanishes at the equator) |

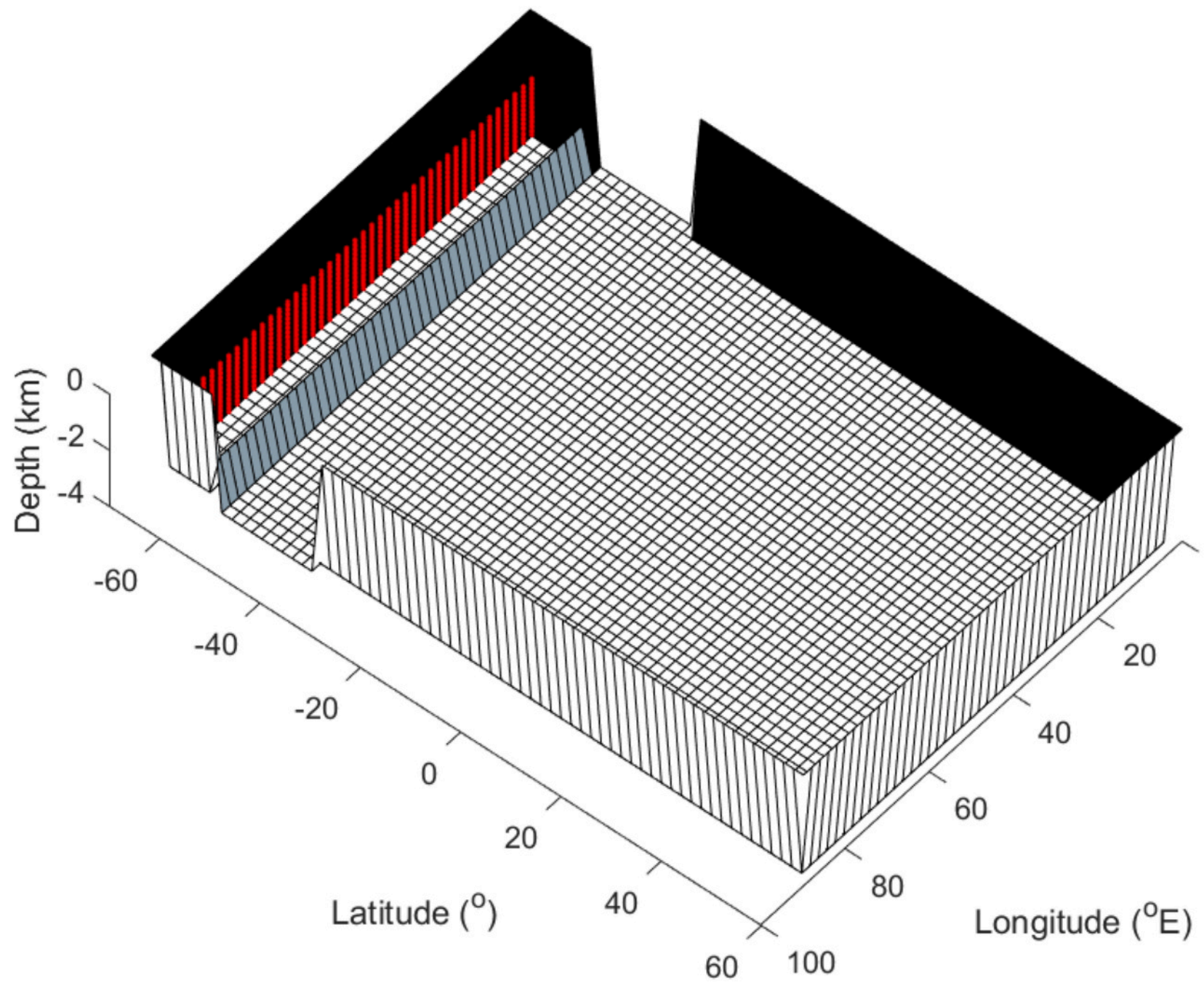

**Fig. 8.** Topography of the model domain and area of forcing. The domain includes an AABW formation region spanning 70°–60°S, a reentrant channel between 60°–40°S (Southern Ocean), and an interior basin extending from 40°S to 60°N (Indo-Pacific). The bathymetry is uniform at a depth of 4000 m, except for a zonal seamount approximately 1500 m high located at 60°S, separating the AABW formation region and the reentrant channel. Red dots mark the southernmost grid points below 2000 m, where temperatures are restored to 1°C on a monthly basis, continuously generating the densest bottom water. The total zonal width of the basin is 92°.

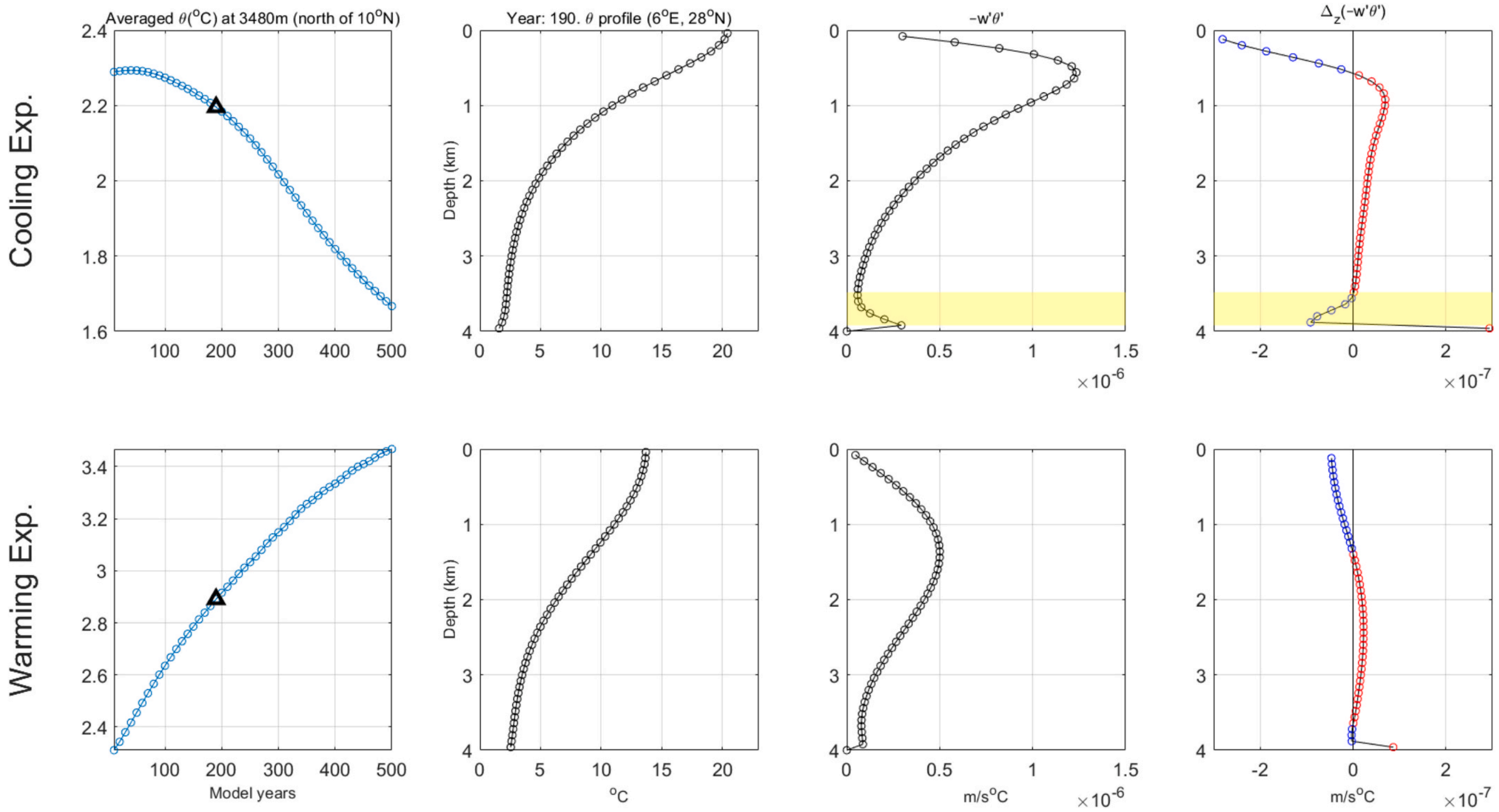

**Fig. 9.** Time series (first column) of the average temperature at a depth of 3,480 m in the interior basin region (10°–60°N) for the cooling (top row) and warming (bottom row) experiments over the first 500 model years. Black triangles mark Year 190, selected for further analysis. At a representative location in the interior basin (6°E, 28°N), vertical profiles of temperature (second column), vertical temperature flux (third column), and vertical convergence of temperature flux (fourth column) are shown for both scenarios at Year 190. The vertical temperature flux vanishes at the seafloor, as demanded by the no-flux boundary condition. Yellow shading highlights the layer of bottom-intensified vertical flux. Blue (red) circles in the fourth column indicate divergence (convergence) of the vertical fluxes.

Thus, warming and cooling in the transient experiments are not caused by surface heat-flux anomalies associated with climate variability. Rather, they are stratification adjustments arising from the balance between AABW supply and mixing-induced transformation. No surface heat-flux forcing is imposed in the transient experiments, allowing us to focus specifically on these abyssal processes.

### *4.2. Simulation results*

#### *4.2.1. Transient solutions: Reproduction of bottom-intensified buoyancy flux*

The cooling and warming trends in the abyss of the interior basin are effectively reproduced in the transient-state configurations, as illustrated in the first column of Fig. 9. From the warming/cooling time series, the model output from the 190$^{th}$ year is selected for analysis, as it most clearly demonstrates the phenomenon under investigation. As will be shown later, the large-scale circulations in both experiments, including the circumpolar gyre and the abyssal overturning circulation, are well spun up by Year 190.

The barotropic streamfunctions indicate that a robust circumpolar gyre circulation develops in both scenarios, despite the absence of wind forcing (Fig. 10). This suggests that thermodynamic forcing, specifically the formation of AABW, can sustain a significant circumpolar gyre. This result aligns with previous modeling studies suggesting that the buoyancy forcing can influence the strength of the Antarctic Circumpolar Current (e.g., Hogg 2010; Shi et al., 2021; Sohail et al., 2025). While the strength of the gyre varies over time in both scenarios, its mean structure is well established by Year 190 (see Year-350 output in Fig. S2 ).

In terms of the abyssal overturning circulations, the residual MOC streamfunction is used for illustration. The residual MOC is defined as the net effect of transport by the Eulerian-mean and eddy-induced circulation (e.g., Marshall and Radko, 2003; Nurser and Lee, 2004; Ito and Marshall, 2008; Munday et al., 2013). The Eulerian and eddy-induced MOC streamfunctions are calculated with the Eulerian velocity and the parameterized bolus velocity, respectively (Gent and McWilliams, 1990; Ito and Marshall, 2008). Both Eulerian and residual MOC streamfunctions are advective overturning diagnostics. In fact, the residual MOC closely resembles the traditional Eulerian MOC in most of the interior basin (Ito and Marshall, 2008), as will be demonstrated later.

It is intriguing to note that the two abyssal cells, as depicted by the residual MOC streamfunctions at Year 190, exhibit upwelling of similar strength in the abyssal interior basin, regardless of whether the system is warming or cooling (Fig. 11, left column). In the cooling experiment, the model diagnostics in Fig. 11 show that isopycnals rise more rapidly than the water parcels, resulting in diagnosed downward diapycnal transport within the bottom 500 m above the seafloor. This model-diagnosed relationship is consistent with the kinematic interpretation illustrated schematically in Fig. 7a and the diagnosed diapycnal downwelling coincides vertically with the bottom-intensified buoyancy-flux layer highlighted in Fig. 9.

In contrast, isopycnals in the warming abyss descend (Fig. 11, lower row), consistent with increasing temperatures or decreasing densities. The diagnostics show descending isopycnals together with rising water parcels. Water therefore moves upward across the isopycnals, producing diagnosed upward diapycnal transport that is stronger than the advective transport because the fluid and isopycnals move in opposite

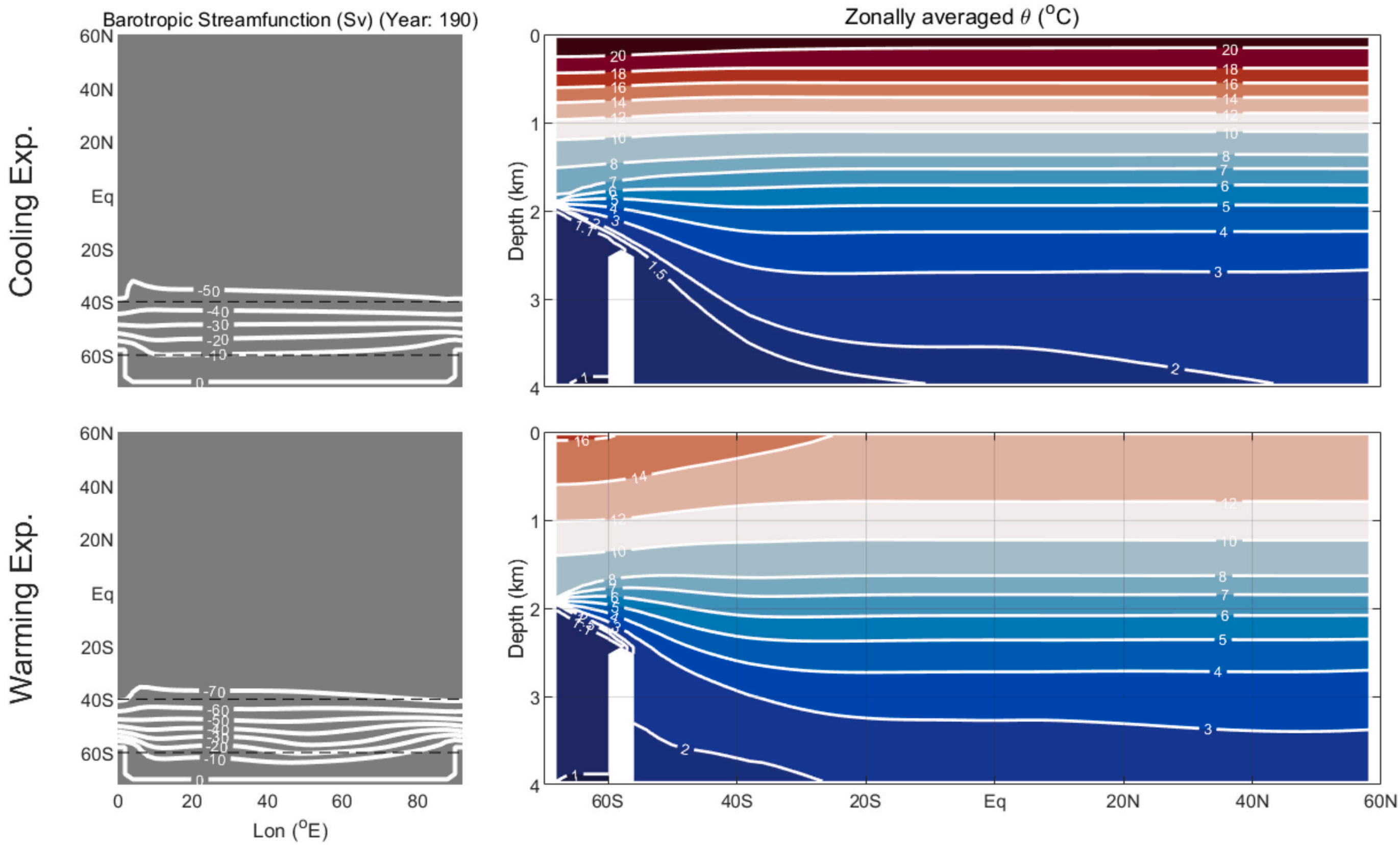

**Fig. 10.** The barotropic streamfunctions (left column) and zonally-averaged potential temperature (right column) for the initial cooling experiment (top row) and warming experiment (bottom row) at Year 190.

directions. Fig. 7c schematically illustrates this model-diagnosed relationship.

An additional feature appears in the upper ocean of the warming experiment (Fig. 11, lower row): pronounced diapycnal downwelling occurs within the upper ~1300 m, accompanied by minimal vertical movement of water parcels. This diapycnal downwelling cell is clearly associated with the upward shifting of isopycnals in the same region (Fig. 11, lower-middle panel). The result offers a further example that helps elucidate the interior-downwelling conundrum raised in Munk's abyssal recipes: that diagnosed diapycnal motion does not necessarily coincide with the vertical trajectory of water parcels during transient adjustment.

For reference, the corresponding overturning diagnostics at Year 350 are provided in Fig. S3. Their overall patterns and abyssal-cell strengths resemble those at Year 190 (Fig. 11), although the downward diapycnal cell in the cooling experiment is substantially weaker. This change indicates that water parcels now ascend in a more adiabatic manner—that is, the advective upwelling is dominated by what Monkman and Jansen (2024) identified as "transient volume tendencies associated with cooling (in the Indo-Pacific)".

While the residual MOC streamfunction ($\psi_{res}$) is employed here for illustration, as mentioned earlier, it closely mirrors the Eulerian MOC streamfunction ($\psi_{Eul}$) in the interior basin (north of the reentrant channel), which is the primary region of interest. The difference between the two, represented by the bolus MOC streamfunction associated with eddies, is significant only south of 20°S, where the abyssal isopycnals exhibit greater tilting (Fig. 10, right panels). For reference, Fig. S4 compares $\psi_{Eul}$ and $\psi_{res}$ in the two scenarios.

A notable outcome of the cooling experiment is that it dynamically generates a bottom-intensified diffusive flux rather than prescribing its vertical profile. The vertical temperature flux is defined as

$$\overline{w'\theta'} = -\kappa\frac{\Delta\theta}{\Delta z}, \tag{9}$$

can be computed using the constant diffusivity and temperature profile obtained from the model output. Its vertical profile compares well with the direct diagnostics of the heat budget terms from MITgcm ('DFrI_TH'+'DFrE_TH' in the model output). Given the ocean's stable stratification, $\overline{w'\theta'}$ remains non-positive throughout. Since a linear equation of state is applied in our simulation (Table 2), the vertical density flux is proportional to the negative temperature flux, i.e., $\overline{w'\rho'} \propto -\overline{w'\theta'}$. Consequently, a bottom-intensified vertical density flux is observed within ~500 m above the BBL in the cooling scenario (Fig. 9, third panel of top row, highlighted by the yellow-shaded band). The e-folding length scale of this bottom-intensified layer is around 231 m ($R^2$=0.97), comparable to the typical vertical e-folding length scale of 500 m (Kunze et al., 2006; McDougall and Ferrari, 2017). This consistency is notable, given the highly idealized nature of our model configuration.

The convergence of the temperature flux ($-\Delta_z\overline{w'\theta'}$) is calculated as the vertical difference of $-\overline{w'\theta'}$, as shown in the fourth column of Fig. 9. Negative (positive) convergence of heat flux corresponds to cooling (warming). In particular, the bottom-intensified vertical flux coincides with the diagnosed diapycnal downwelling cell over the same depth range (Fig. 11, upper-right panel). This bottom-intensification structure is pervasive within the abyssal ocean of the cooling experiment. Supporting evidence is found in the widespread occurrence of negative vertical heat convergence in the profile below 3000 meters at nearly all grid points in the interior basin (not shown). These results underscore that, even under simplified external forcing, the cooling experiment reproduces the observed bottom-intensified mixing.

#### 4.2.2. *Quasi-equilibrium solution*

To demonstrate the MOC behavior under approximately steady conditions, a separate experiment was conducted. Building upon the cooling configuration, an additional SST relaxation was imposed over the equatorial region (3°S-3°N) to introduce a heat source that counteracts the polar heat sink. This adjustment enables the system to approach equilibrium and conceptually mirrors the SST-relaxation approach used by FMM16. The model was integrated for 4,000 years to reach quasi-equilibrium.

The results are presented in Fig. 12. As anticipated, a stable upwelling MOC cell dominates the abyssal ocean. The isopycnal volume tendency ($\psi_{iso}$), diagnosed from isopycnal displacements, is essentially zero under steady stratification, leading to a near-perfect agreement between the structures of the advective overturning and kinematically diagnosed diapycnal transport.

#### 4.2.3. *An additional transient cooling solution*

The cooling experiment presented in Section 4.2.1 shows that the advective overturning ($\psi_{res}$) and isopycnal volume tendency ($\psi_{iso}$) are both directed upward, whereas the diagnosed diapycnal transport ($\psi_{dia}$) is directed downward. Because the diagnosed diapycnal transport is defined kinematically as $\psi_{dia} = \psi_{res} - \psi_{iso}$, the direction of $\psi_{dia}$ depends on the relative magnitudes of the two upward components. This relationship therefore permits another cooling configuration in which the advective overturning, $\psi_{res}$, exceeds the isopycnal volume tendency, $\psi_{iso}$, in magnitude, resulting in upward rather than downward diagnosed diapycnal transport.

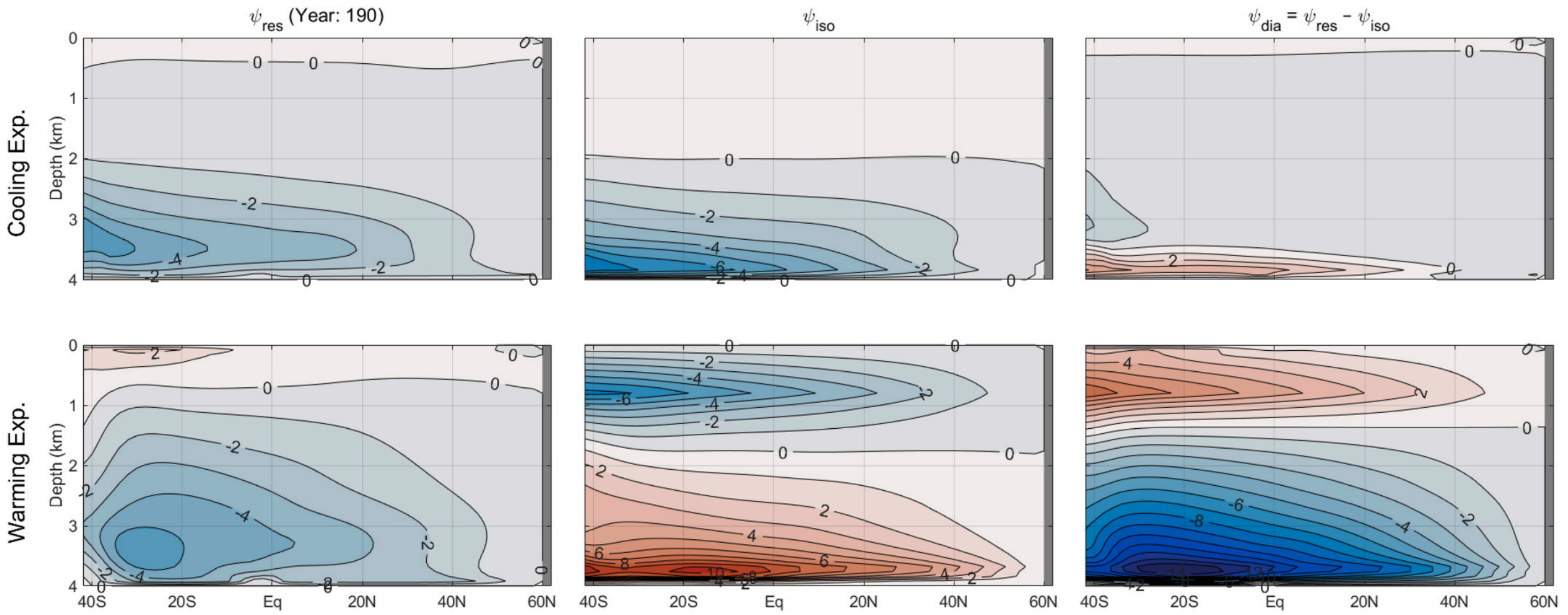

**Fig. 11.** The three overturning diagnostics at Year 190 for the initial cooling experiment (upper row) and the warming experiment (lower row): residual MOC streamfunction ($\psi_{res}$; left column), isopycnal volume tendency ($\psi_{iso}$; middle column), and kinematically diagnosed diapycnal transport ($\psi_{dia}$; right column). Positive (negative) values or warm (cold) colors represent clockwise (anticlockwise) overturning cells. Units: Sv. Contour interval: 1 Sv.

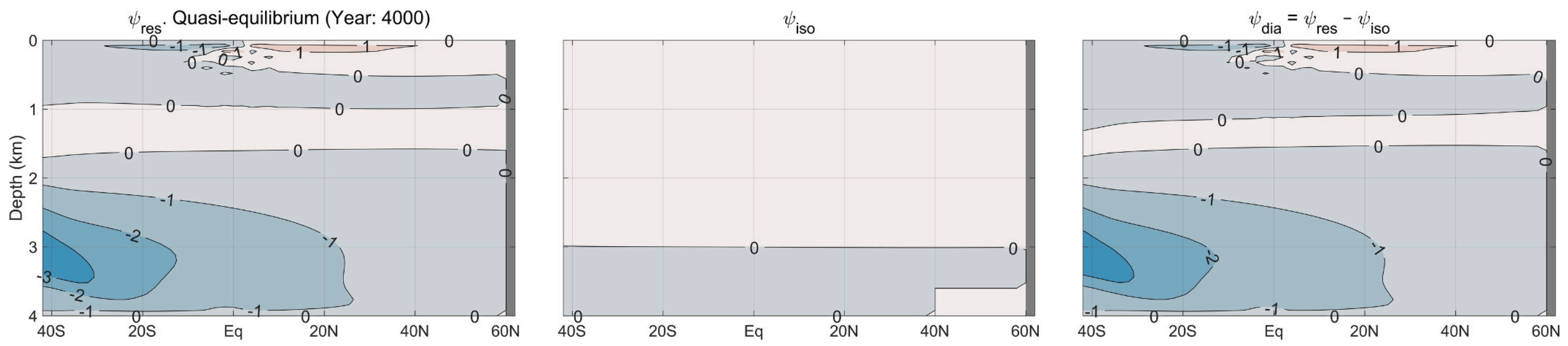

**Fig. 12.** As in Fig. 11, but based on a simulation integrated over 4,000 years to reach quasi-equilibrium.

Such a configuration may occur if mixing-induced water-mass transformation is stronger than in the original cooling experiment. Stronger mixing would transform dense water more rapidly, thereby reducing the rate of abyssal cooling and the associated isopycnal volume tendency. However, provided that mixing remains insufficient to transform dense water faster than it is supplied by AABW intrusion, the abyss should continue to cool. This reasoning suggests that the alternative cooling configuration may be realized using a diffusivity larger than that in the original cooling experiment but smaller than that in the warming experiment.

Motivated by this possibility, we conducted an additional experiment with an intermediate interior-basin vertical diffusivity of $3 \times 10^{-5}$ $m^2/s$, between the values used in the original cooling ($1 \times 10^{-5}$ $m^2/s$) and warming ($1 \times 10^{-4}$ $m^2/s$) experiments. All other model parameters remain unchanged.

The numerical diagnostics are shown in Fig. 13. The continued upward displacement of isopycnals indicates that the abyss remains in a transient cooling state. Between 2,500 and 3,500 m, the advective overturning transport and isopycnal volume tendency are both directed upward. Unlike in the original cooling experiment, however, the upward advective overturning exceeds the isopycnal volume tendency in magnitude. Consequently, the diagnosed diapycnal transport is also directed upward. Thus, all three diagnostics exhibit an upward sense within this depth range.

The two cooling solutions are therefore kinematically distinct and cannot be represented by a single overturning relationship. We hereafter refer to the original cooling solution in Section 4.2.1 as Type-I cooling (Fig. 11) and the additional cooling solution in Fig. 13 as Type-II cooling. In both regimes, the advective overturning and isopycnal volume tendency are directed upward. In Type-I cooling, $|\psi_{iso}| > |\psi_{res}|$, yielding downward diagnosed diapycnal transport. In Type-II cooling, $|\psi_{iso}| < |\psi_{res}|$, yielding upward diagnosed diapycnal transport. The limiting condition $|\psi_{iso}| = |\psi_{res}|$ corresponds to zero diagnosed diapycnal transport, i.e., an ideally adiabatic upwelling. Fig. 7a depicts the Type-I cooling relationship. Although the Type-II relationship is not shown separately, it can be visualized by interchanging the final relative positions of the water parcel and isopycnal $\rho_0$, such that the water parcel rises faster than the isopycnal and the diagnosed diapycnal velocity reverses upward.

## 4.3. Discussion

The numerical diagnostics in Figs. 11–13 reveal two kinematically distinct cooling regimes, in addition to the warming and quasi-equilibrium regimes. In Type-I cooling, both the advective overturning and isopycnal volume tendency are directed upward in the near-bottom layer, yielding downward diagnosed diapycnal transport. This downward transport occurs within the same near-bottom region as the bottom-intensified buoyancy flux diagnosed in Fig. 9. In Type-II cooling, within the 2,500–3,500 m depth range, all three diagnostics are directed upward. Thus, transient abyssal cooling can be associated with either upward or downward diagnosed diapycnal transport.

In the warming experiment, rising water parcels coexist with descending isopycnals, and the diagnosed diapycnal transport is upward and stronger in magnitude than the advective overturning. In the quasi-equilibrium experiment, the isopycnal volume tendency approaches zero as the stratification becomes statistically stationary. The advective and diagnosed diapycnal transports therefore become nearly identical, consistent with the steady-stratification assumption underlying the classical "abyssal recipes" framework.

To examine whether these relationships depend on the choice of vertical coordinate, we also calculated the three overturning diagnostics in potential-temperature coordinates for all four experiments (Fig. S5). Because salinity is held constant in the model and a linear equation of state is used, potential temperature provides a monotonic proxy for potential density. The large-scale patterns and relationships are qualitatively consistent with those obtained in depth coordinates. One additional feature appears in Type-I cooling: a diapycnal upwelling cell occurs within the coldest, densest water classes, consistent with the ongoing transformation of the densest bottom water toward lighter classes. This cell may be less apparent in depth coordinates because fixed-depth integration combines contributions from different water classes.

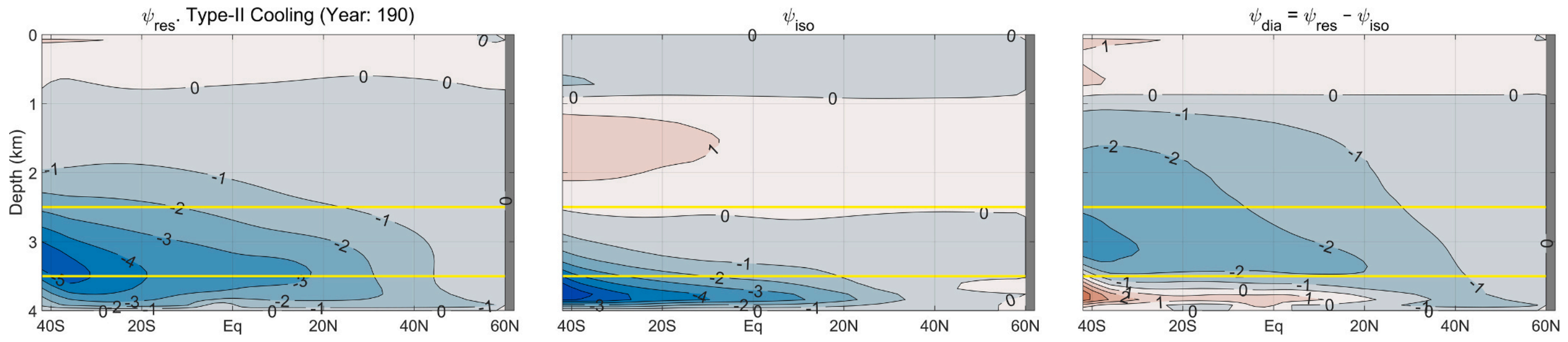

**Fig. 13.** As in Fig. 11, but for the additional transient cooling experiment with an intermediate interior-basin vertical diffusivity of $3 \times 10^{-5}$ $m^2/s$. The yellow lines mark depths of 2,500 and 3,500 m. Within this range, the advective overturning transport, isopycnal volume tendency, and kinematically diagnosed diapycnal transport are all directed upward. This solution is referred to as Type-II cooling in the text.

The relationships consistently diagnosed in both depth and potential-temperature coordinates are summarized schematically in Fig. 14. The schematic provides a conceptual synthesis rather than independent evidence for these relationships. Together, the numerical diagnostics show that the transient thermal state does not uniquely determine the diagnosed diapycnal response; its direction is determined kinematically by the relative magnitudes and directions of the advective overturning and isopycnal volume tendency.

The Type-I experiment provides a transient interpretation of the interior-downwelling conundrum: advective upwelling can coexist with downward kinematically diagnosed diapycnal transport when the upward isopycnal volume tendency exceeds the advective transport in magnitude. The Type-II experiment further shows that such downwelling is not an inevitable consequence of abyssal cooling. Realistic bathymetry, spatially heterogeneous mixing, wind forcing, and thermohaline variations may modify these relationships and determine which regime is realized in the ocean. Indeed, Meza and Gebbie, 2025 demonstrated that wind stress adjustments can enhance residual upwelling and isopycnal shoaling in the North Pacific, highlighting a dynamical forcing pathway deliberately excluded from the present idealized experiments.

## 5. Summary

This study investigates how abyssal overturning circulation behaves under transient and steady-state conditions. The motivation stems from Munk's abyssal recipes, which implicitly assume a steady-state abyssal ocean. However, accumulating evidence suggests that the contemporary abyssal ocean, particularly in the Indo-Pacific, is not in equilibrium and instead exhibits widespread changes in stratification. These ongoing changes raise the question of how the relationship between advective overturning and diapycnal transport under transient conditions may differ from the steady-state relationship assumed in classical theories.

To examine this question, we developed four idealized numerical experiments driven by AABW formation. Unlike some previous idealized configurations, abyssal isopycnals do not outcrop at the surface and no vertical-buoyancy-flux profile is prescribed, allowing warming and cooling to emerge dynamically through the interaction between AABW formation, mixing, and evolving stratification.

Within the idealized model framework examined here, advective upwelling occurs in all four experiments, whereas the diagnosed diapycnal response differs among them. In the near-bottom layer of Type-I cooling, the upward isopycnal volume tendency exceeds the advective overturning in magnitude, yielding downward diagnosed diapycnal transport. In Type-II cooling, the advective overturning is stronger than the isopycnal volume tendency within the 2,500–3,500 m depth range, so all three diagnostics are directed upward. The warming and quasi-equilibrium experiments also exhibit upward diagnosed diapycnal transport, although for different reasons: the isopycnal volume tendency opposes the advective overturning during warming and approaches zero under quasi-equilibrium. Thus, the Type-I cooling illustrates that advective upwelling can coexist with downward kinematically diagnosed diapycnal transport, whereas Type-II cooling shows that Eulerian cooling alone does not determine the direction of this diagnostic.

The Type-I cooling experiment spontaneously produces an observationally documented bottom-intensified buoyancy-flux profile, with a simulated e-folding length scale comparable to previous estimates reported in the literature. It also produces overturning structures qualitatively consistent with those diagnosed from the ECCO cooling realization of the abyssal Indo-Pacific. These independent comparisons suggest that, despite its highly idealized configuration, the model may provide a useful conceptual reference for interpreting transient abyssal overturning.

The numerical configuration builds on the box-basin experiment of FMM16, although the objectives and forcing strategies differ. Type-I cooling produces basin-interior residual upwelling together with downward kinematically diagnosed diapycnal transport, resembling the FMM16 solution obtained with an imposed bottom-intensified buoyancy-flux profile. Here, however, this overturning structure arises due to transient stratification adjustment, and the bottom-intensified profile emerges dynamically. Type-II cooling further demonstrates that downward kinematically diagnosed diapycnal transport is not an inevitable consequence of abyssal cooling. The experiments therefore provide a complementary transient kinematic interpretation on the interior-downwelling conundrum addressed by FMM16 under steady-state conditions.

Overall, this study represents an initial exploration of abyssal overturning dynamics under transient conditions. The principal finding is that, within the idealized model examined here, advective abyssal upwelling can coexist with either upward or downward kinematically diagnosed diapycnal transport during transient cooling. The direction of this diagnostic depends on the relative magnitudes and directions of advective overturning and isopycnal-volume adjustment. These relationships remain qualitatively consistent in potential-temperature coordinates. However, this coordinate consistency does not establish a general equivalence between the kinematic diagnostic and a thermodynamic water-mass-transformation rate; such an equivalence is neither assumed nor demonstrated here. The results therefore provide a transient kinematic interpretation of the interior-downwelling conundrum, with its applicability beyond the present idealized framework requiring further investigation.

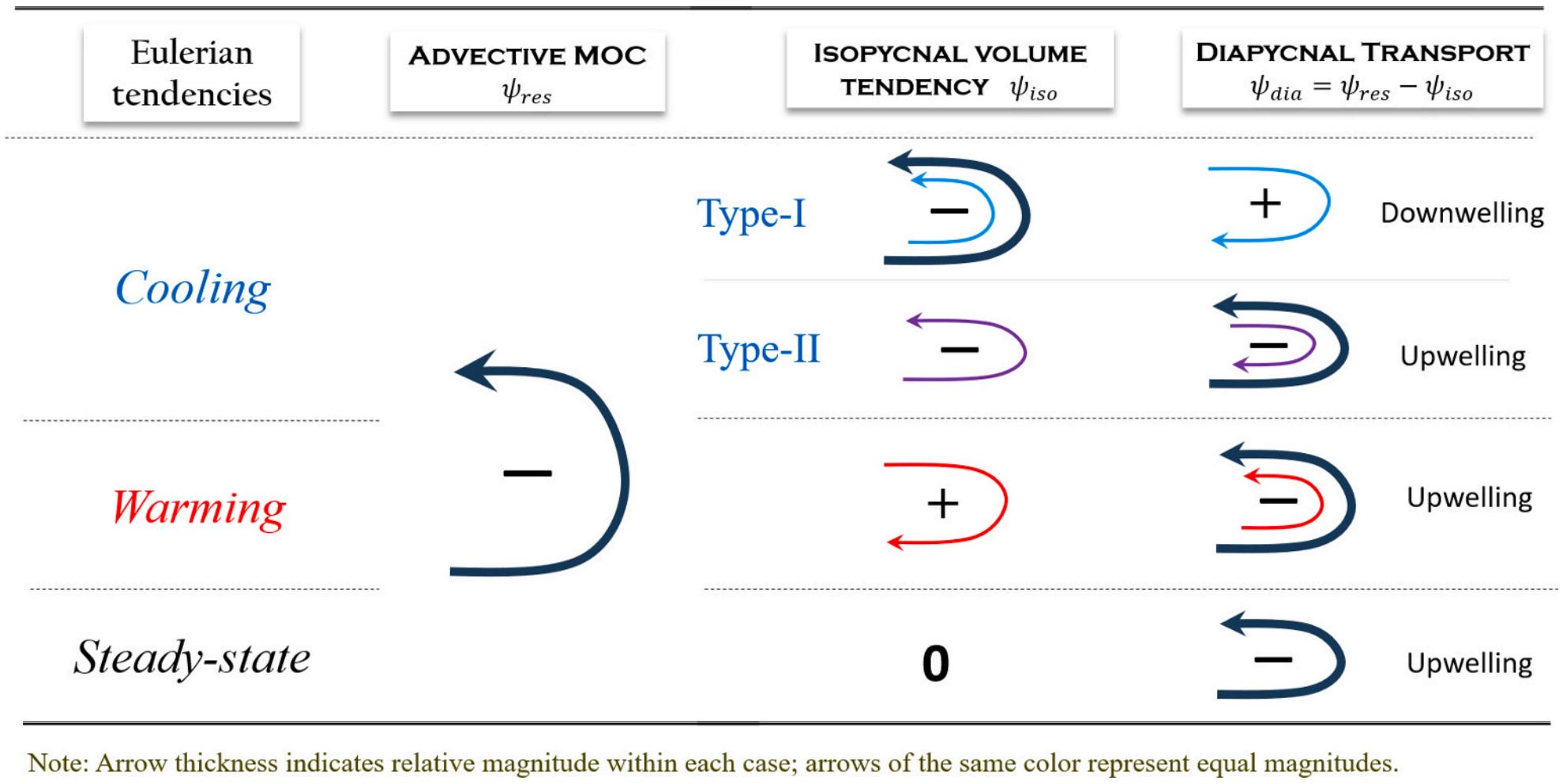

**Fig. 14.** Schematic synthesis of the relationships among the advective (residual) overturning circulation, isopycnal volume tendency ($\psi_{iso}$), and kinematically diagnosed diapycnal transport ($\psi_{dia} = \psi_{res} - \psi_{iso}$) under (a) Type-I and Type-II cooling, (b) warming, and (c) equilibrium conditions. Negative (positive) streamfunction values represent anti-clockwise (clockwise) overturning cells. Arrow thickness indicates relative magnitude within each case; arrows of the same color represent equal magnitudes. The schematic summarizes the numerical results presented in Figs. 11–13 and Fig. S5.

Several important questions remain open. Because the present experiments intentionally simplify wind forcing, bathymetry, salinity variations, and the spatial structure of mixing, their implications should be interpreted within the idealized framework examined here. By holding salinity constant, the experiments isolate density adjustment expressed through temperature variations. In the real ocean, salinity variations may modify both the density response and AABW source properties, potentially affecting the simulated transient overturning. Further modeling studies are needed to evaluate the robustness and broader applicability of the results. Potential avenues include:

- *Imposing wind forcing* above the reentrant channel to investigate the respective roles of dynamical and thermodynamic forcing in driving the abyssal overturning circulations in the interior basin.
- *Reactivating SST restoration* to study the interplay between the upper and lower overturning cells.
- *Varying the intensity of interior mixing and rate of AABW formation* to better understand the respective roles of buoyancy-related "pulling" and "pushing" in sustaining the abyssal overturning cells.
- *Incorporating geothermal heat flux* to evaluate its influence on the strength and structure of abyssal overturning.
- *Incorporating more realistic bathymetry, salinity variations, and spatially heterogeneous mixing* to evaluate the sensitivity of the simulated transient behavior to these processes.

## Availability statement

The MITgcm model code is available at https://mitgcm.readthedocs.io/en. The specific modeling configuration and output files used in this study have been archived on Zenodo and can be accessed via https://zenodo.org/records/15640739. The ECCO reanalysis dataset (Version 4, Release 3) is openly available at https://ecco.jpl.nasa.gov.

## CRediT authorship contribution statement

**Lei Han:** Writing – review & editing, Writing – original draft, Visualization, Validation, Supervision, Software, Resources, Project administration, Methodology, Investigation, Funding acquisition, Formal analysis, Data curation, Conceptualization.

## Declaration of competing interest

The author declares that they have no known competing financial interests or personal relationships that could have appeared to influence the work reported in this paper.

## Acknowledgments

This work was supported by NSFC (grants 42476006 and 42676002) and the Xiamen University Malaysia Research Fund (grant XMUMRF/2024-C14/ICAM/0017). The author thanks the reviewers for their careful evaluation and constructive comments, which greatly improved the clarity and rigor of the manuscript.

## Appendix A. Supplementary data

Supplementary data to this article can be found online at https://doi.org/10.1016/j.pocean.2026.103850.